\documentclass[sigconf, nonacm]{acmart}
\usepackage{amsmath}
\usepackage{tabularx}
\usepackage{booktabs}
\usepackage{enumitem}
\usepackage{multirow}
\let\classAND\AND
\let\classOR\OR
\let\classXOR\XOR
\let\classNOT\NOT

\let\AND\relax
\let\OR\relax
\let\XOR\relax
\let\NOT\relax

\usepackage{algorithm}
\usepackage{algorithmic}

\let\AND\classAND
\let\OR\classOR
\let\XOR\classXOR
\let\NOT\classNOT

\AtBeginEnvironment{algorithmic}{\let\AND\algoAND}
\AtBeginEnvironment{algorithmic}{\let\OR\algoOR}
\AtBeginEnvironment{algorithmic}{\let\XOR\algoXOR}
\AtBeginEnvironment{algorithmic}{\let\NOT\algoNOT}
\usepackage [
n, % or lambda advantage , operators , sets,
adversary , landau , probability , notions ,
logic ,
ff, mm,
primitives , events , complexity , oracles , asymptotics , keys
]{ cryptocode}

\usepackage{mdframed}

\newtheorem{gadget}{\bfseries Gadget}

\newcommand{\reducespaceThalf}{\vspace{-2mm}}
\newcommand{\reducespaceT}{\vspace{-4mm}}

\begin{document}


\title[VPAS: Publicly Verifiable and Privacy-Preserving Aggregate Statistics on Distributed Datasets]{VPAS: Publicly Verifiable and Privacy-Preserving Aggregate Statistics on Distributed Datasets}

%%%%%%%%%%%%%%%% Authors' Info %%%%%%%%%%%%%%%%%

\author{Mohammed Alghazwi}
\affiliation{%
 \institution{University of Groningen}
 \city{Groningen}
 \country{Netherlands}}
 \email{m.a.alghazwi@rug.nl}
\orcid{0000-0003-4039-4748 }

\author{Dewi Davies-Batista}
\affiliation{%
 \institution{University of Groningen}
 \city{Groningen}
 \country{Netherlands}}
 \email{d.j.davies-batista@student.rug.nl}

\author{Dimka Karastoyanova}
\affiliation{%
 \institution{University of Groningen}
 \city{Groningen}
 \country{Netherlands}}
 \email{d.karastoyanova@rug.nl}
\orcid{0000-0002-8827-2590}

\author{Fatih Turkmen}
\affiliation{%
 \institution{University of Groningen}
 \city{Groningen}
 \country{Netherlands}}
 \email{f.turkmen@rug.nl}
\orcid{0000-0002-6262-4869}


\renewcommand{\shortauthors}{Alghazwi et al.}

\begin{abstract}

Aggregate statistics play an important role in extracting meaningful insights from distributed data while preserving privacy. A growing number of application domains, such as healthcare, utilize these statistics in advancing research and improving patient care. 

In this work, we explore the challenge of input validation and public verifiability within privacy-preserving aggregation protocols. We address the scenario in which a party receives data from multiple sources and must verify the validity of the input and correctness of the computations over this data to third parties, such as auditors, while ensuring input data privacy. To achieve this, we propose the "VPAS" protocol, which satisfies these requirements. Our protocol utilizes homomorphic encryption for data privacy, and employs Zero-Knowledge Proofs (ZKP) and a blockchain system for input validation and public verifiability. We constructed VPAS by extending existing verifiable encryption schemes into secure protocols that enable $N$ clients to encrypt, aggregate, and subsequently release the final result to a collector in a verifiable manner.

We implemented and experimentally evaluated VPAS with regard to encryption costs, proof generation, and verification. The findings indicate that the overhead associated with verifiability in our protocol is $10\times$ lower than that incurred by simply using conventional zkSNARKs. This enhanced efficiency makes it feasible to apply input validation with public verifiability across a wider range of applications or use cases that can tolerate moderate computational overhead associated with proof generation.

\end{abstract}

\begin{CCSXML}
<ccs2012>
   <concept>
       <concept_id>10002978.10002991.10002995</concept_id>
       <concept_desc>Security and privacy~Privacy-preserving protocols</concept_desc>
       <concept_significance>500</concept_significance>
       </concept>
 </ccs2012>
\end{CCSXML}

\ccsdesc[500]{Security and privacy~Privacy-preserving protocols}

\keywords{Privacy, Homomorphic Encryption, Zero-Knowledge Proofs, Blockchain, Genomics, Privacy-preserving protocols}

\maketitle

\section{Introduction} \label{sec:intro}
Aggregate statistics play a crucial role in deriving meaningful insights from large, often geographically distributed datasets. The core concept of an aggregation system is straightforward: given a dataset \(x_1, \ldots, x_n\) from clients \(U_1, \ldots, U_n\), the aggregator executes a function \(f(x_1, \ldots, x_n)\) and forwards the outcome to a collector $C$ for analysis \cite{corrigan2017prio, davis2023verifiable}.

Traditional methods for calculating these statistics involve transmitting sensitive data in plaintext to aggregator servers for central processing, raising significant privacy concerns and potential for misuse. This is particularly concerning as the amount of sensitive data being collected and processed continues to grow in various domains, such as healthcare and finance.

To mitigate privacy risks while enabling valuable statistical analysis of sensitive data, there has been a shift towards decentralized computing augmented by privacy-enhancing technologies (PETs), leading to the development of various privacy-preserving data aggregation methods. There are two core issues VPAS addresses in this context.

\textbf{Input Validation.}
A crucial aspect in deploying secure aggregation in practice is correctness and input data validation in the face of corrupted clients. This involves augmenting the aggregation protocols with safeguards against clients attempting to manipulate the results by inserting malicious or malformed input. Moving from a scenario where participants are assumed to be honest but curious to one where they may act maliciously can be accomplished through the application of zero-knowledge proofs, as indicated in various proposals \cite{froelicher2020drynx,corrigan2017prio,davis2023verifiable}. These prior approaches focus on input range and validating correct encoding. In particular, range proofs can be used to prove in zero-knowledge that an input value is within a given range. Although effective, these methods are confined to verifying that inputs fall within a certain range. Our work addresses the challenge of employing general-purpose zero-knowledge proofs for broader validation tasks.

\textbf{Public verifiability.}
Ensuring verifiability, particularly public verifiability, is another crucial aspect when aggregate statistics are used in critical domains such as healthcare \cite{malghazwi2022blockchain,bontekoe2023verifiable}. Public verifiability allows any party, including non-participants, to confirm the correctness of the computed results according to the protocol's specifications without accessing the underlying data. By applying public verifiability to the computation, we can eliminate the need to re-run the computation since we rely on cryptographic methods to publicly verify that given the input and the functions, the same results will always be obtained. This feature not only ensures result reproducibility but also facilitates audit processes by documenting data usage, computational functions, and the distribution of results, thereby promoting responsible data sharing. 

\textbf{ Contributions.}
In this work, we introduce a novel, lightweight protocol for aggregate statistics, named VPAS (Publicly \textbf{V}erifiable \textbf{P}rivacy-preserving \textbf{A}ggregate \textbf{S}tatistics), that employs homomorphic encryption and zero-knowledge proofs among others in order to support input validation, computational correctness, and public verifiability. VPAS supports a wide range of aggregation tasks, such as sum, mean, standard deviation, and linear regression.

A key contribution of VPAS over prior works is the \emph{distributed verifiable encryption} (DVE), a new technique that enables verifiable encryption to operate in distributed settings. Complemented by verifiable aggregation and re-encryption mechanisms, our approach significantly advances the prior works by enabling a broad range of input validation tasks through general-purpose Zero-Knowledge Succinct Non-Interactive Argument of Knowledge (zkSNARK). 
%We also optimize the performance of the zkSNARK scheme used in VPAS. 
Our protocol offers robust security without relying on non-colluding servers and ensures that proofs of computation correctness are publicly verifiable.

We demonstrate the practicality of VPAS through a prototype implementation and a detailed case study on private aggregate statistics for distributed genomic data. Our evaluation reveals that VPAS provides higher performance when compared to simply using conventional zkSNARKs, offering a tenfold ($10\times$) increase in client performance.

\subsection{Application Scenario: GWAS} \label{subsec:app-scenario}
Privacy-preserving aggregation systems such as VPAS can be applied in various settings. In this section, we briefly discuss an application scenario from which many of the design requirements for VPAS have been drawn (more details in Appendix~\ref{gwas}).

\textbf{Aggregate statistics on distributed genomic data.}
In genomics research, Genome-Wide Association Studies (GWAS) involve computing aggregate summary statistics on highly personal genome data. GWAS facilitates the identification of genes associated with specific diseases, potentially leading to targeted treatments and personalized medicine. However, achieving statistically significant results requires access to a large number of individual genotypes. The sensitive nature of genetic data often results in individuals' reluctance to share their data for research purposes \cite{clayton_systematic_2018}, posing a substantial challenge. The beacon framework \cite{fiume2019federated} plays a pivotal role in enhancing the accessibility of genomic data, enabling researchers to locate datasets containing specific genetic variants of interest. Recently, there has been an advancement in the beacon framework to support more complex computations on aggregate genome data \cite{rambla2022beacon}. This enhanced beacon framework can provide aggregated results, such as the count of individuals carrying a specific variant, which is invaluable for epidemiological research. The primary challenge lies in utilizing Privacy-Enhancing Technologies (PETs) to maintain privacy while ensuring the scientific research's essential properties of correctness, authenticity, and reproducibility are upheld. With the integration of a suitable input validation circuit to verify the authenticity of inputs, VPAS can support these crucial properties, thereby serving as a significant enhancement to the beacon framework. To demonstrate this, we have configured and evaluated VPAS for this application scenario as will be discussed later in section \ref{sec:eval}.

\section{Related Work}\label{sec:related-work}
Traditional approaches for privacy-preserving computation of aggregate statistics primarily employ centralized systems \cite{popa2011cryptdb,liu2012mona} using cryptographic methods or trusted hardware \cite{bogdanov2015students}. These solutions often rely on a single trusted entity, which in turn poses a single point of failure. Our distributed setting is similar to recent privacy-preserving protocols \cite{bater2016smcql,jagadeesh2017deriving,melis2015efficient,bogdanov2008sharemind,gascon2016privacy,yang2018private}, yet these often omit input validation and public verifiability, crucial for secure data analysis in distributed contexts.

There are also some works that focus input data validation for aggregate statistics. Yang and Li \cite{yang2013detecting} developed a protocol identifying out-of-range values via re-encryption. However, this method is particularly susceptible to collusion attacks, where collusion between the aggregator and a user is enough to compromise the privacy of all group members by disclosing their values to the aggregator. Protocols like UnLynx \cite{froelicher2017unlynx}, Drynx \cite{froelicher2020drynx}, and ACORN \cite{bell2023acorn} improve the efficiency of input validation but are limited to range checks, making them less generic to other validation tasks.  

Secret-shared Non-Interactive Proof (SNIPs), introduced in Prio by Corrigan-Gibbs and Boneh \cite{corrigan2017prio}, allow for arbitrary input validation but demand multiple honest (non-colluding) servers. Additionally, the communication costs grow linearly with the complexity of the validation circuit which renders this approach impractical when the validation circuit is large. Prio3 \cite{davis2023verifiable} and Elsa \cite{rathee2023elsa} extend these ideas with optimizations yet still depend on non-colluding servers. In contrast to these methodologies, our protocol facilitates arbitrary input validation and public verifiability without assuming server non-collusion. To the best of our knowledge, no existing protocol encompasses all these attributes. A comparative summary is presented in Table \ref{tab:comp}.

Given that aggregate statistics on distributed genomic datasets serve as a use case and are used in our evaluation, we now discuss the related work in this domain. In the genomic literature, extensive research focuses on data privacy, with initiatives like the annual IDASH competition\footnote{\url{http://www.humangenomeprivacy.org/}} promoting privacy-focused solutions. Yet, there is a notable gap in research focusing on the verifiability or validity of inputs in the context of privacy-preserving data aggregation within genomics. Recent efforts \cite{halimi2022privacy, jiang2023reproducibilityoriented} address result verification using differential privacy. These works highlight the importance of verifiability and reproducibility in genomic research as we do in this work. However, it is essential to recognize that the scope of differential privacy in the context of verifying genome-wide association study (GWAS) results diverges significantly from that of Zero-Knowledge Proofs (ZKPs). While the former is concerned with the verification of published findings, it does not address the secure and private aggregation of data nor the assurance of correctness and public verifiability, which are central to our study. Furthermore, ZKPs offer a more substantial guarantee by affirming the accuracy of the output relative to valid inputs and the specified computational process, thereby providing a more comprehensive and verifiable assurance of the output's accuracy.

\begin{table}[ht]
\centering
\small
\begin{tabular}{cccccc}
\hline
\textbf{Work} & \textbf{Crypto.} & \textbf{Data} & \textbf{Comput.} & \textbf{Public} & \textbf{Input} \\
& \textbf{scheme} & \textbf{Privacy} & \textbf{Correct.} & \textbf{Verif.} & \textbf{Valid.} \\
\hline
\\
This work & HE+ZKP & \checkmark & \checkmark & \checkmark &\checkmark\\ \\
Yang et.al. \cite{yang2013detecting} & HE+ZKP  & \checkmark & x & x & x \\ \\
Unlynx \cite{froelicher2017unlynx} & HE+ZKP  & \checkmark & \checkmark & \checkmark & x \\ \\
Drynx \cite{froelicher2020drynx} & HE+ZKP & \checkmark & \checkmark & \checkmark & x \\ \\
ACORN \cite{bell2023acorn} & HE+ZKP  & \checkmark & \checkmark & x & x \\ \\
Elsa \cite{rathee2023elsa} & MPC+ZKP & \checkmark & x & x & x \\ \\
Prio \cite{corrigan2017prio} & MPC+ZKP  & \checkmark & x & x & \checkmark \\ \\
Prio3 \cite{davis2023verifiable} & MPC+ZKP  & \checkmark & x & x & \checkmark \\ \\
\hline
\end{tabular}
\caption{Comparative analysis of various systems assuming our system and threat model. 'Input Valid.' refers to protocols that allow for arbitrary input validation.}
\label{tab:comp}
\end{table}

\section{Preliminaries} \label{sec:prelim}
In the following, we briefly describe the main building blocks and cryptographic schemes that we used to build our system. 

\subsection{Threshold (Partial) Homomorphic Encryption} \label{subsec:phe-elgamal}
A partially homomorphic encryption (HE) scheme is a public key encryption scheme that allows limited computation over the ciphertexts. The HE scheme we consider in this work is the exponential ElGamal encryption which is an additive homomorphic encryption scheme \cite{cramer1997secure}. A threshold variant of this scheme (t out of n) has some additional properties. While the public key is known to everyone, the secret key $\sk$ is split across a set of clients $U=$$\{U_1,\dots,U_n\}$ such that a subset of them ($S \subseteq U$) must participate together to reconstruct the secret key. The threshold structure can be altered based on the adversarial assumption. In this work, we use a threshold structure where all parties must participate (i.e., $t=n=|S|=|U|$) in order to decrypt a ciphertext. In the following, we describe the ElGamal encryption scheme appropriate to our setting. It supports distributed key generation and re-encryption.  

Let us consider $n$ clients $U = \{U_1, \ldots, U_n\}$ and a designated party $C$, called as the collector (more details about the participants in Section~\ref{subsec:sys-model}). The encryption scheme operates over a cyclic group $G$ of order ${\mid}G{\mid}$ with a generator $g$. Each client $U_i$ independently generates its private key $\sk_i$ by selecting a value uniformly at random from $\{1, \ldots, {\mid}G{\mid} - 1\}$ and then computes the public key as $\pk_i = g^{\sk_i}$. The public keys of all clients are aggregated to create the collective-key $\pk_{\alpha}$. The distributed key generation algorithm can be described as:
\begin{equation*}
\text{DKG:} ({1^\lambda}) \rightarrow (\pk_{\alpha} = \prod_{i=1}^{n} \pk_i = \prod_{i=1}^{n} g^{\sk_i})
\end{equation*}

Each client can then encrypt their input data using $\pk_{\alpha}$ by splitting the data into messages with a maximum bit length of $k$, where $2^k \leq {\mid}G{\mid}$. The encryption function of a message $m$ is the following, where $r \in \{0, . . . , {\mid}G{\mid} - 1\}$ is a uniformly chosen randomness:
\begin{equation*}
Encrypt: (m, \pk_\alpha, r) \rightarrow (c_1= g^r, c_2=g^m{\cdot}{\pk_\alpha^r})
\end{equation*}

Since exponential ElGamal encryption is additively homomorphic, the following function illustrates the addition of two ciphertexts $(c_1, c_2) \oplus (d_1, d_2) := (c_1 \cdot d_1, c_2 \cdot d_2)$:
\begin{equation*}
\begin{split}
Add: ((c_1,c_2),(d_1,d_2)) \rightarrow (c_1 \cdot d_1 = g^{r_1+r_2}, \\ 
c_2 \cdot d_2 = g^{m_1+m_2}\cdot{pk_\alpha^{r_1+r_2}})
\end{split}
\end{equation*}

In our work, the ciphertexts resulting from aggregation are only released to the collector, therefore, a re-encryption protocol such as the one described in \cite{froelicher2020drynx} can be used for this purpose. Each client partially decrypts the ciphertext, i.e., removes the effect of their public key \(\pk_i\), and re-encrypts it with the collector's public key \(\pk_\beta\). The results will be an encrypted message under \(\pk_\beta\) only if all clients participate in the protocol. The re-encryption protocol is described as follows:

Given a ciphertext \((C_1, C_2)\) encrypted with $\pk_{\alpha}$, each client computes \(w_1^i\) and \(w_2^i\) using a secret uniformly-random value \(z_i\):
\begin{equation*}
(w_1^i, w_2^i) \leftarrow (w_1^i = g^{z_i}, w_2^i = C_1^{|G| - \sk_i} \cdot \pk_{\beta}^{z_i})
\end{equation*}

Then, each pair \((w_1^i, w_2^i)\) is combined to form the re-encrypted ciphertext:
\begin{equation*}
(\hat{C_1}, \hat{C_2}) \leftarrow (\hat{C_1} = \prod_{i=1}^{n} w_1^i, \hat{C_2} = C_2 \cdot \prod_{i=1}^{n} w_2^i)
\end{equation*}

The decryption function of a ciphertext $(\hat{C_1}, \hat{C_2})$ using the private key $\sk_\beta$ to obtain the message $m$ is the following:
\begin{equation*}
Decrypt: (\sk_\beta,(\hat{C_1}, \hat{C_2})) \rightarrow (m={log_g(\hat{C_2}{\cdot}\hat{C_1}^{{\mid}G{\mid}-\sk_\beta})})
\end{equation*}
As seen above, the decryption function requires solving a discrete logarithm $log_g(x)$ where $g$ is the base.

\subsection{Zero Knowledge Proofs (ZKPs)} \label{subsec:zkp}
Zero-knowledge proof schemes \cite{blum2019non} provide mechanisms for a prover to prove the knowledge of a secret to a verifier with overwhelming probability, without revealing the secret itself. Zero-Knowledge Succinct Non-Interactive Argument of Knowledge (zkSNARK) \cite{bitansky2012extractable} is arguably the most popular ZKP scheme and also the one we employ in our work. Specifically, we use the Groth16 scheme \cite{groth2016size}. A zkSNARK scheme involves the following algorithms:

\begin{itemize}
[leftmargin=0.4cm]
    \item $\text{CRS} \leftarrow \text{Setup}(\mathcal{R})$: takes an arbitrary relation $\mathcal{R}$ as an input, and outputs the corresponding common reference string CRS. 

    \item $\pi \leftarrow \text{Prove}(CRS,\Phi,\omega)$. Given the $CRS$, a statement $\Phi$, and a witness $\omega$ such that ($\Phi, \omega$) $\in \mathcal{R}$, generate a proof $\pi$.

    \item $0/1 \leftarrow \text{Verify}(CRS,\Phi,\pi)$. Given the $CRS$, a statement $\Phi$, and a proof $\pi$, output 1 if the proof is valid and 0 otherwise. 
\end{itemize}

\subsection{Verifiable Encryption} \label{subsec:ve}
A verifiable encryption scheme is one in which it is possible to prove specific properties of a message $M$, given only its encryption $CT$. This ensures that the plaintext is encrypted correctly and satisfies the criteria written in a general-purpose zkSNARK circuit. In this work, we focus on the SAVER framework (SNARK-friendly, Additively-homomorphic, Verifiable Encryption and Decryption with Rerandomization)~\cite{lee2019saver}. SAVER has been applied to various domains, including digital payments \cite{bontekoe2022balancing}. Since we only require verifiable encryption in our work, we provide the definition of SAVER's verifiable encryption omitting the decryption and rerandomization components. For any arbitrary zkSNARK relation $\mathcal{R}$, SAVER's verifiable encryption scheme comprises the following four algorithms:
\begin{itemize}
[leftmargin=0.5cm]
    \item $\text{CRS} \leftarrow \text{Setup}(\mathcal{R})$: takes an arbitrary relation $\mathcal{R}$ as an input, and outputs the corresponding common reference string CRS.
    \item $\sk, \pk \leftarrow \text{KeyGen}(\text{CRS})$: takes a CRS as an input, and outputs a secret key $\sk$, and a public key $\pk$.
    \item $\pi, \text{CT} \leftarrow \text{Enc}(\text{CRS}, ~$\pk$, M, \Phi, \omega)$: takes CRS, a public key $\pk$, a message $M = \{m_1, \ldots, m_n\}$, a statement $\Phi = \{\phi_{n+1}, \ldots, \phi_l\}$, and a witness $\omega$ as inputs, and generates a proof $\pi$ and a ciphertext CT = $(c_0, \ldots, c_n, \psi)$.
    \item $0/1 \leftarrow \text{VerifyEnc}(\text{CRS}, \pi, \text{CT}, \Phi)$: takes CRS, a proof $\pi$, a ciphertext CT, and a statement $\Phi = \{\phi_{n+1}, \ldots, \phi_l\}$ as inputs, and outputs 1 if (CT, $\Phi$) $\in$ $\mathcal{R}$, or 0 otherwise.
\end{itemize}

\subsection{Blockchains and Distributed Ledgers}
VPAS relies on a distributed ledger for storing proofs. We employ blockchain, which we summarize here, in implementing this ledger however any bulletin-like platform can also be used. Blockchain stores and verifies transactions on a ledger that is distributed to all nodes in a peer-to-peer (P2P) network. The transactions are organized into blocks that are protected by a combination of cryptographic techniques to ensure the integrity of the recorded transactions. A consensus protocol is then followed to validate the blocks and the blocks that are successfully validated are added to the growing chain of blocks. This essentially solves the problem of allowing multiple parties that do not necessarily trust each other to agree on the state of a shared ledger. Throughout the remainder of this paper, the term 'distributed ledger' will be used to denote this concept more generally.

\section{System Overview} \label{sec:overview}

\subsection{System Model} \label{subsec:sys-model}
\begin{figure}[h]
  \includegraphics[width=0.4\textwidth]{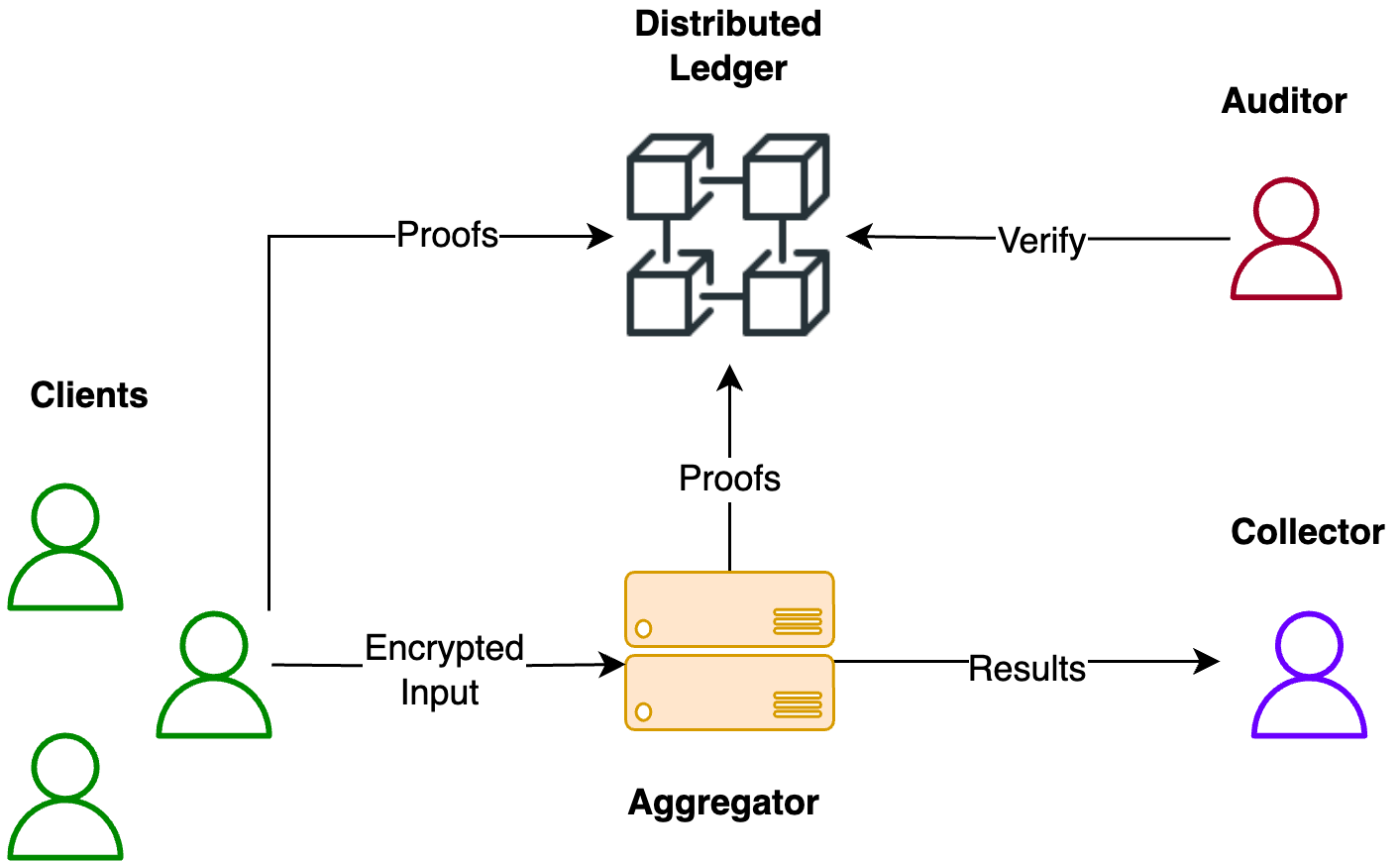}
  \caption{VPAS System Overview showing the components of the system and their interactions. The clients send their encrypted data to the aggregator and proofs to the distributed ledger. The aggregator processes the input, sends the result to the collector, and submits proof to the distributed ledger. The auditor verifies the execution of the protocol by inspecting the distributed ledger.\label{fig:overview}}
\end{figure}

We illustrate the system components, participants, and their interactions in Fig. \ref{fig:overview} and describe them in the following. Conceptually, our system model resembles the model of existing architectures for aggregate statistics such as the ones outlined in prior works \cite{corrigan2017prio, froelicher2020drynx}. In particular, we assume a federation of multiple data owners, each having access to private data, and a collector requesting statistical analysis on the combined datasets from the federation. Therefore, there are three main participants: the collector, the clients, and the aggregator. Additionally, since we consider public verifiability, we add the auditor and distributed ledger as participants in the system. We describe all system participants and components in the following:

\begin{itemize}
[leftmargin=0.4cm]
    \item \emph{Collector:} The entity submitting a query for specific statistics. If accepted the collector will receive the final aggregate results of the analysis.
    \item \emph{Clients:} Owners or custodians of the private datasets. Each client is entrusted with information and authorized to use this data. Clients can be organizations that maintain databases on their premises and can perform operations over this data locally, or they could be individuals that own a single data point. 
    \item \emph{Aggregator:} Entity that collects all the input data from the data owners, and computes the aggregation function. 
    \item \emph{Distributed ledger:} A ledger that records query executions and ensures data availability, tamper-proofing, and auditability. This ledger maintains a transparent and immutable record of all operations, providing traceability and verifiability.
    \item \emph{Auditor:} Wants to check the correctness of the query execution. The auditor has access to the distributed ledger which includes the necessary parameters for verification. The auditor can be the collector, the clients, or an independent entity monitoring the execution. 
\end{itemize}

\vspace{-0.5cm}

\subsection{Security Goals} \label{subsec:sec-goals}
Our goal with VPAS is to design a lightweight, publicly verifiable, privacy-preserving aggregation protocol with input validation that meets the following criteria:

\emph{(a) Privacy: } The system is viewed as private if no party participating in the protocol can learn anything about the inputs beyond their own. Additionally, the output of the protocol should only be revealed to the collector. The collector learns nothing about the input $x_1, \cdots , x_n$ beyond what it can infer from the aggregate result $y = f(x_1, \cdots , x_n)$.

\emph{(b) Robustness: } The outcome of the aggregation is robust if malicious clients are unable to compromise the aggregate result by submitting malformed or inaccurate inputs. This definition aligns with that provided in previous studies \cite{corrigan2017prio}. However, whereas prior research was mainly focused on enforcing input range, our work addresses more complex scenarios where client input must be validated using a general-purpose zkSNARKs circuit. An example of such validation includes proving that the submitted input is part of a public Merkle tree commitment.

\emph{(c) Correctness: } Ensuring the accurate execution of the computations done on the client input during all phases of the protocol to guarantee the correctness of the aggregation result. The correctness guarantee is essential since we assume the system participants might be malicious. 

\emph{(d) Public Verifiability:} The protocol should enable public verification of its execution. This feature allows all entities, including those not directly involved in the protocol, to verify or audit its execution.

\subsection{Threat Model} \label{subsec:threat-model}
We make the assumption that all parties involved in the protocol have reached a consensus on a specific aggregation function to be computed and have also agreed to share the final result of the computation with the collector. We consider the \emph{anytrust} threat model similar to that used in \cite{froelicher2020drynx, wolinsky2012scalable}. This model operates under the assumption that all parties, except one, may potentially be compromised by a malicious attacker. This implies that the compromised parties have the ability to deviate from the protocol in various ways, such as providing inconsistent inputs, substituting their own input with that of another party, or performing computations different from the expected ones. Additionally, a malicious client may collude with any other party in the system including the aggregator and collector to try to infer information about individual data. 

\textbf{Out-of-scope threats}
In this work, we do not consider attacks carried out by the collector on the aggregation result. Exploring techniques to prevent the results from revealing excessive details about the data is a complementary direction, and we anticipate that some of these techniques can be integrated into the system. For example, combining our approach with differential privacy tools that introduce noise before the release of the results can help ensure that the results do not leak excessive information about individual records in the input dataset.  

\section{VPAS Construction} \label{sec:construction}
To construct VPAS, we need to incorporate mechanisms for data protection and input validation. For the former, we rely on ElGamal encryption described in section \ref{subsec:phe-elgamal}, and for the latter we employ zkSNARKs. In order to combine the two, we constructed the following cryptographic gadgets, which are used in the VPAS protocol (described in section \ref{sec:protocol}) to meet the requirement for input validation, correctness, and public verifiability. We define these gadgets as follows:

% \vspace{1em}
% Gadget 1: Distributed Verifiable encryption (DVE)
\begin{gadget}[Distributed Verifiable Encryption (DVE)]
A gadget whereby the encrypted data is accompanied by a proof that assures the validation of publicly-defined properties. This gadget can be summarized by the following algorithms:
\begin{itemize} [leftmargin=0.4cm]
    \item $CRS \leftarrow \text{Setup}(\mathcal{R})$
    Takes an arbitrary relation $\mathcal{R}$ as an input, and outputs the corresponding common reference string (CRS).
    
    \item
    $pk_\alpha \leftarrow \text{DKG}(\text{CRS})$
    Takes a CRS as an input, performs a distributed key generation protocol, and outputs the collective public key $pk_\alpha$.
    
    \item 
    $\pi, CT \leftarrow \text{Enc}(\text{CRS}, pk_\alpha, M, \Phi, \omega)$: takes CRS, a public key pk, a message $M = \{m_1, \ldots, m_n\}$, a statement $\Phi = \{\phi_{n+1}, \ldots, \phi_l\}$, and a witness $\omega$ as inputs, and generates a proof $\pi$ and a ciphertext CT = $(c_0, \ldots, c_n, \psi)$.
    
    \item 
    $0/1 \leftarrow \text{Verify Enc}(CRS, \pi, CT, \Phi)$
    takes CRS, a proof $\pi$, a ciphertext CT, and a statement $\Phi = \{\phi_{n+1}, \ldots, \phi_l\}$ as inputs, and outputs 1 if (CT, $\Phi$) $\in$ $\mathcal{R}$, or 0 otherwise.
\end{itemize}
\end{gadget}

% Gadget 2: Verifiable Aggregation (VA)
\begin{gadget}[Verifiable Aggregation (VA)]
A gadget that, when presented with a set of ciphertexts $\{CT_1, \cdots, CT_n\}$, not only generates the combined result using a designated aggregation function \(f\), but also proves the accuracy of this aggregation procedure. This gadget includes the following algorithms:

\begin{itemize} [leftmargin=0.4cm]
    \item
    $\pi, \hat{CT} \leftarrow \text{Agg}(\{CT_i\}_{i=0}^{n})$
    Given the ciphertexts \(\{CT_1, \cdots ,CT_n\}\), outputs a proof of correct aggregation \(\pi\) and the resulting ciphertext \(\hat{CT}\).
    
    \item
    $0/1 \leftarrow \text{VerifyAgg}(\pi, \hat{CT}, \{CT_i\}_{i=0}^{n})$
    takes a proof \(\pi\), ciphertexts \(\hat{CT}\) and \(\{CT_i\}_{i=0}^{n}\) as inputs, and outputs 1 if \(\hat{CT}\) correctly aggregates \(\{CT_i\}_{i=0}^{n}\), or 0 otherwise.
\end{itemize}

\end{gadget}

% Gadget 3: Verifiable Re-Encryption (VRE)
\begin{gadget}[Verifiable Re-Encryption (VRE)]
A gadget designed to validate the conversion of a ciphertext \( CT \), initially encrypted using a public key \( pk_\alpha \), into another ciphertext \( \hat{CT} \) encrypted with \( pk_\beta \). This gadget ensures that the underlying plaintext value remains unchanged throughout this conversion process, and without decryption at any stage. As described in the re-encryption protocol in Section~\ref{subsec:phe-elgamal}, this process requires participation from all clients, with each client $U_i$ needing to produce re-encryption shares, denoted as \( (w_1^i, w_2^i) \). The gadget comprises the following algorithms:

\begin{itemize} [leftmargin=0.4cm]
    \item 
    $\{\pi, w_1^i, w_2^i\} \leftarrow \text{GenShare}(CT, pk_\beta, pk_i, sk_i, z_i)$
    Given a ciphertext \(CT = (C_1, C_2)\), a new public key \(pk_\beta\), the keypair $(pk_i, sk_i)$ used by client $U_i$, and a random value $z_i$, output a proof \(\pi\) and the re-encryption shares \((w_1^i, w_2^i)\).
    
    \item 
    $\hat{CT} \leftarrow \text{ReEnc}((CT, \{w_1^i, w_2^i, \pi_i\}_{i=1}^n, pk_\beta, \{pk_i\}_{i=1}^n))$
    
    Takes a re-encryption proof \(\pi\), the shares \((w_1^i, w_2^i)\) from clients $U_1, \cdots, U_n$, a ciphertext $CT$, the public keys of each client $pk_i$, and the new public key \(pk_\beta\) as inputs, and outputs $\hat{CT}$ if all \((w_1^i, w_2^i)\) shares are the correct re-encryption shares, or aborts otherwise.
\end{itemize}

\end{gadget}

\subsection{Construction I} \label{subsec:const1}
A straightforward approach to achieve this is to include the entire encryption algorithm in the zk-SNARK circuit along with the generic relation to ensure the consistency of m between $Prove$ and $Enc$. We refer to this approach as \emph{encryption-in-the-circuit}.

The input validation and the entire encryption process are integrated within the zkSNARKs circuit. Unfortunately, Such an approach faces efficiency issues when dealing with complex cryptographic operations. For instance, additively homomorphic encryption such as Paillier encryption imposes a significant computational burden on the prover, rendering it impractical for large datasets \cite{visscher2022poster}. We address this issue by using the Elgamal encryption which introduces minimal overhead compared to other HE schemes. 

For the experimental evaluation, we implemented the DVE, VA, VRE gadgets in the zkSNARK circuit. Additionally, we optimized the implementation to ensure that the required cryptographic operations are performed at the lowest possible cost. Details of our optimized techniques and implementation are elaborated upon in Appendix~\ref{ecc-opt}.

\subsection{Construction II} \label{subsec:const2}
In this section, we demonstrate our construction that separates the encryption from the zkSNARKs circuit using an extended version of SAVER's verifiable encryption \cite{lee2019saver}. The verifiable encryption algorithms in SAVER are not suitable for our model, primarily due to the construction of the public key. In SAVER, it is assumed that a single party generates the public key $pk$, and private key $sk$, as demonstrated in \cite{lee2019saver}. However, in a distributed setting with multiple collaborating parties, as in our model, this approach does not sufficiently ensure privacy for the raw datasets unless a trusted party is involved. Since our threat model does not assume the presence of a trusted party, we replace the SAVER key generation algorithm with a distributed variant. Additionally, we exclude other SAVER algorithms as they are irrelevant to our setting, and replace them with additional functionalities, namely, verifiable aggregation (VA), and verifiable re-encryption (VRE). We specify the core algorithms of construction II in Algorithm \ref{alg:dve}, \ref{alg:va}, and \ref{alg:vre}. 

\emph{Distributed Verifiable Encryption (DVE).}
We have tailored the verifiable encryption scheme to the distributed setting that we consider in this work. Specifically, our algorithm diverges from \cite{lee2019saver} in the manner in which the encryption key is generated and subsequently utilized for encryption and verification. Our modified key generation algorithm (DKG) in Algorithm~\ref{alg:dve} accounts for the fact that the private key is distributed among $k$ number of clients ($U_1, ..., U_k$). The partial key generation produces the secret key $sk_j$ and the partial public key $pk_j$ for client $U_j$. The secret key contains two sets of random values, $\{s_i\}_{i=1}^n$ and $\{t_i\}_{i=0}^n$. Using these values, the partial public key ($pk_j$) is constructed. The partial public key $pk_j$ omits the $P_1$ value. This omission is necessary since $P_1$'s generation requires combining all clients' partial keys, particularly the $X = \{G^{\delta s_i}\}_{i=1}^n$ component of $pk_j$. The clients broadcast their partial keys which are then merged to form an aggregated partial public key $\hat{pk}$. Subsequently, each client computes a share of the $P_1$ value, which is then broadcasted and aggregated to produce the final $P_1$ value. The collective public key $pk_\alpha$ is formed by merging $\hat{pk}$ with the aggregated $P_1$ value. This $pk_\alpha$ is then utilized for encryption and verification, analogous to SAVER's Algorithm~\ref{alg:dve}. Note that in the distributed key generation, the verification key VK, used in SAVER, is discarded as it is not relevant to our setting.

\emph{Verifiable Aggregation (VA).}
The ciphertexts produced by the DVE algorithm are additively homomorphic, functioning as outlined in Algorithm~\ref{alg:va}. Given that the input ciphertexts have already been submitted as part of the Distributed Verifiable Encryption (DVE) proof, the aggregator needs to only supply the resulting ciphertexts. These can then be verified for correctness against the input ciphertexts. 

\emph{Verifiable Re-Encryption (VRE).}
The re-encryption shares, denoted as $(w_1^j, w_2^j)$ for client $U_j$, are generated from the ciphertext $CT$, as outlined in Algorithm~\ref{alg:vre}. Additionally, a Proof of Knowledge of the discrete logarithm (PoK) is constructed. This PoK is used to verify that the re-encryption shares $(w_1^j, w_2^j)$ have been correctly computed. The proof is publicly verifiable and can be verified using the client's public key, $pk_j$, the collector's public key $pk_\beta$, and the ciphertext, $CT$. Our work adapts the proofs of knowledge of discrete logarithms as described in \cite{camenisch1997proof} to work with the ciphertexts generated from the DVE scheme described earlier. To prove the correctness of the re-encryption shares $(w_1^j, w_2^j)$, the prover proves that two discrete logarithms satisfy a linear equation. In our case, given that $pk_j = (X_0, X^j, Y^j, Z^j, P_2^j)$, the discrete logarithms of ($X^j, w_1^j$), should satisfy the linear equation:
\[
-C_1^{s} \cdot pk_\beta^{z} = w_2^j
\]
The public values are $C_1, pk_\beta, pk_j, w_1^j, w_2^j$, and the private values are ($s, z$). Essentially the verifier checks that the $s$ and $z$ values used in ($pk_j, w_1^j$) are equal to the ones used to generate $w_2^j$. We elaborate further in Appendix \ref{pok}.

\begin{algorithm}
\caption{Distributed Verifiable Encryption (DVE)}
\label{alg:dve}

% setup
\begin{algorithmic}
\STATE
\STATE \textbf{Setup}($\mathcal{R}$)
\end{algorithmic}
\begin{algorithmic}[1]
\STATE $\mathcal{R}$($m_1, \ldots, m_n, \phi_{n+1}, \ldots, \phi_l; \omega$):
\STATE $\hat{CRS} \leftarrow \Pi_{snark}\text{.Setup}(\mathcal{R})$
\STATE $\text{CRS} \leftarrow \hat{CRS} \cup \{G^{-\gamma}\}$
\end{algorithmic}
\begin{algorithmic}
\STATE \textbf{return} $CRS$
\end{algorithmic}

% Distributed keygen
\begin{algorithmic}
\STATE
\STATE \textbf{DKG}($CRS$)
\end{algorithmic}
\begin{algorithmic}[1]
\STATE \textbf{Partial key generation:}
\FOR{$j = U_1, U_2, \ldots, U_k$}
\STATE $\{s_i\}_{i=1}^n, \{t_i\}_{i=0}^n \stackrel\$\leftarrow \mathbb{Z}^*_p$
\STATE $sk_j \leftarrow (\{s_i\}_{i=1}^n, \{t_i\}_{i=0}^n)$
\STATE $pk_j \leftarrow \left(G^\delta,\{G^{\delta s_i}\}_{i=1}^n, \{G^{t_i}\}_{i=1}^n, \{H^{t_i}\}_{i=0}^n, G^{-\gamma \sum_{i=1}^n s_i}\right)$
\ENDFOR
\STATE \textbf{Combine partial keys:}
\STATE \text{let} $pk_j = (X_0, X^j, Y^j, Z^j, P_2^j)$
\STATE $\hat{pk} \leftarrow (X_0, \prod_{j=1}^k X^j, \prod_{j=1}^k Y^j, \prod_{j=1}^k Z^j, G^{-\gamma} \cdot \prod_{j=1}^k P_2^j)$
\STATE \textbf{Generate P1 shares:}
\FOR{$j = U_1, U_2, \ldots, U_k$}
\STATE parse $sk_j = (\{s_i\}_{i=1}^n, \{t_i\}_{i=0}^n)$
\STATE let $\hat{pk} = (X_0,\{X_i\}_{i=1}^n,\{Y_i\}_{i=1}^n,\{Z_i\}_{i=0}^n,P_2)$
\STATE $P_1^j = X_0^{t_0} \cdot \prod_{i=1}^n X_i^{t_i}$
\ENDFOR
\STATE \textbf{Combine P1 shares:}
\STATE $P_1 = \prod_{j=1}^k P_1^j$
\STATE \textbf{Generate $pk_\alpha$:}
\STATE let $\hat{pk} = (X_0,X,Y,Z,P_2)$
\STATE $pk_\alpha = (X_0,X,Y,Z,P_1,P_2)$
\end{algorithmic}
\begin{algorithmic}
\STATE \textbf{return} $(\{pk_j\}_{j=1}^k, \{sk_j\}_{j=1}^k, pk_\alpha)$
\STATE
\end{algorithmic}

% enc
\begin{algorithmic}
\STATE \textbf{Enc}($CRS, pk_\alpha, m_1, \ldots, m_n, \phi_{n+1}, \ldots, \phi_l; \omega$)
\end{algorithmic}
\begin{algorithmic}[1]
\STATE \text{let} $pk_\alpha = (X_0, \{X_i\}_{i=1}^n, \{Y_i\}_{i=1}^n, \{Z_i\}_{i=0}^n, P_1, P_2)$
\STATE $r \stackrel\$\leftarrow \mathbb{Z}^*_p$
\STATE $\text{CT} \leftarrow (X_0^r, X_1^r G_1^{m_1}, \ldots, X_n^r G_n^{m_n}, \psi = P_1^r \cdot \prod_{i=1}^n Y_i^{m_i})$
\STATE $\hat{\pi} \leftarrow \Pi_{snark}\text{.Prove}(\text{CRS}, m_1, \ldots, m_n, \phi_{n+1}, \ldots, \phi_l; \omega)$
\STATE $\pi \leftarrow (A, B, C \cdot P_2^r)$
\end{algorithmic}
\begin{algorithmic}
\STATE \textbf{return} $(\pi, CT)$
\STATE
\end{algorithmic}

% ver enc
\begin{algorithmic}
\STATE \textbf{VerifyEnc}($CRS, pk_\alpha, \pi, CT, \phi_{n+1}, \ldots, \phi_l$)
\end{algorithmic}
\begin{algorithmic}[1]
\STATE \text{parse} $\pi = (A, B, C)$ and $\text{CT} = (c_0, \ldots, c_n, \psi)$
\STATE \text{let} $pk_\alpha = (X_0, \{X_i\}_{i=1}^n, \{Y_i\}_{i=1}^n, \{Z_i\}_{i=0}^n, P_1, P_2)$
\STATE \text{assert} $\prod^n_{i=0} e(c_i, Z_i) = e(\psi, H)$
\STATE \text{assert} $e(A, B) = e(G^{\alpha}, H^{\beta}) \cdot e(\prod^n_{i=0} c_i \cdot \prod^l_{i=n+1} G_i^{\phi_i}, H^{\gamma}) \cdot $
\end{algorithmic}
\begin{algorithmic}
\STATE $\hspace{2.3cm} e(C, H^{\delta})$
\STATE \textbf{return} $0/1$

\end{algorithmic}
\end{algorithm}
\vspace{-0.5cm}

\begin{algorithm}
\caption{Ciphertext Aggregation}
\label{alg:va}
\begin{algorithmic}
\STATE \textbf{Agg($CT_1$, $CT_2$)}
\end{algorithmic}
\begin{algorithmic}[1]
\STATE \text{let} $ CT_1 \leftarrow (X_0^{r^1}, \{X_i^{r^1} G_i^{m_i^1}\}_{i=1}^n, P_1^{r^1} \prod_{j=1}^n Y_j^{m_j^1})$
\STATE \text{let} $ CT_2 \leftarrow (X_0^{r^2}, \{X_i^{r^2} G_i^{m_i^2}\}_{i=1}^n, P_1^{r^2} \prod_{j=1}^n Y_j^{m_j^2})$
\STATE $CT  = CT_1 \cdot CT_2$
\STATE $\hspace{0.39cm}$ $= (X_0^{r^1+r^2}, \{X_i^{r^1+r^2} G_i^{m_i^1+m_i^2}\}_{i=1}^n, P_1^{r^1+r^2} \prod_{j=1}^n Y_j^{m_j^1+m_j^2})$
\end{algorithmic}
\begin{algorithmic}
\STATE \textbf{return} $CT$
\end{algorithmic}
\end{algorithm}
\vspace{-0.5cm}

\begin{algorithm}
\caption{Verifiable Re-Encryption (VRE)}
\label{alg:vre}
\begin{algorithmic}
% re-enc generate share
\STATE \textbf{GenShare($CT$, $pk_j$, $sk$, $pk_\beta$)}
\end{algorithmic}
\begin{algorithmic}[1]
\STATE \text{let} $ CT = (C_1, C_2, \psi)$
\STATE \text{parse} $sk = (\{s_i\}_{i=1}^n, \{t_i\}_{i=0}^n)$
\STATE \text{let} $pk_j = (X_0, X, Y, Z, P_2)$
\STATE $z \leftarrow \mathbb{Z}^*_p$
\STATE $w_1  = g^z$ 
\STATE $w_2  = (C_1^{|G| - s_1} \cdot pk_\beta^{z}, \dots,  C_1^{|G| - s_n} \cdot pk_\beta^{z})$
\STATE $\pi \leftarrow \Pi_{PoK}.\text{Prove}(\{s_i\}_{i=1}^n, z, X, pk_\beta, w_1, w_2, C_1)$
\end{algorithmic}
\begin{algorithmic}
\STATE \textbf{return} $({w_1,w_2,\pi})$
\STATE
\end{algorithmic}

\begin{algorithmic}
% re-enc aggregate
\STATE \textbf{ReEnc}$(CT, \{w_1^j, w_2^j, \pi_j\}_{j=1}^n, pk_\beta, \{pk_j\}_{j=1}^n)$
\end{algorithmic}
\begin{algorithmic}[1]
\STATE \text{let} $ CT = (C_1, C_2, \psi)$
\FOR{$j=0$ to $n$}
\STATE assert $\Pi_{PoK}.\text{Verify}(\pi_j,pk_j, pk_\beta, C_1, w_1^j, w_2^j)$
\ENDFOR
\STATE $\hat{C_1} \leftarrow \prod_{j=1}^n w_1^j$
\STATE $\hat{C_2} \leftarrow C_2 \cdot \prod_{j=1}^n w_2^j$
\STATE $\hat{CT} \leftarrow (\hat{C_1}, \hat{C_2})$
\end{algorithmic}
\begin{algorithmic}
\STATE \textbf{return} $\hat{CT}$
\end{algorithmic}
\end{algorithm}
\vspace{-0.5cm}

\section{VPAS Protocol} \label{sec:protocol}
In this section, we describe the VPAS protocol in its entirety, utilizing the gadgets previously defined and constructed as foundational building blocks. The VPAS protocol encompasses four main phases: Setup, Submit, Aggregate, and Release. Each of these phases serves a specific purpose, as outlined in Figure \ref{fig:protocol_overview} and further explained in the following.

\begin{figure}[ht!]
\centering
\begin{mdframed}
Considering Clients $U = \{U_1, \ldots, U_n\}$ with private inputs $x = \{x_1, \ldots, x_n\}$, an aggregation function \(f\), an aggregator (server) \(\mathcal{S}\), a collector \(\mathcal{C}\), an auditor $\mathcal{D}$, and a distributed ledger \(\mathcal{L}\).
\begin{enumerate}
[leftmargin=0.5cm]  
    \item \textbf{Setup:} Clients $U$ collectively generate an encryption key \(pk_\alpha\) and a Common Reference String (CRS) for the zkSNARK circuits; they then publish these to \(\mathcal{L}\).

    \item \textbf{Query:} \(\mathcal{C}\) posts the query \(Q\) to \(\mathcal{L}\), specifying an aggregation function \(f\) and a public key \(pk_\beta\).

    \item \textbf{Submit:} Each client \(U_i\) executes the \emph{DVE.Enc} algorithm to produce an encrypted input \(E(x_i)\) and a proof \(\pi_{DVE}^i\), then sends \(E(x_i)\) to \(\mathcal{S}\) and \(\pi_{DVE}^i\) to \(\mathcal{L}\).

    \item \textbf{Aggregate:} \(\mathcal{S}\) verifies each client's input and employs the \emph{VA.Agg} algorithm to compute the aggregation function \(f\), producing the result \(y\) and a proof \(\pi_{VA}\). \(\mathcal{S}\) then publishes \(\pi_{VA}\) to \(\mathcal{L}\) and sends the aggregated result \(y\) to all clients \(U\).

    \item \textbf{Release:} Each client \(U_i\) verifies the received result \(y\) and generates partial re-encryption shares \((w_1^i, w_2^i)\) along with a proof \(\pi_{\text{VRE}}^{i}\) using the \emph{VRE.GenShare} algorithm. The shares are forwarded to \(\mathcal{C}\), and the proof \(\pi_{\text{VRE}}^{i}\) is submitted to ledger \(\mathcal{L}\). \(\mathcal{C}\) aggregates and verifies the re-encryption shares using \emph{VRE.ReEnc}, derives $\hat{y}$, and decrypts it using the secret key $sk_\beta$.

    \item \textbf{Audit:} $\mathcal{D}$ retrieves the public parameters and proofs from $\mathcal{L}$ and verifies each step of the aggregation by executing the verification algorithm for each gadget.
    
\end{enumerate}
\end{mdframed}
\caption{VPAS Protocol}
\label{fig:protocol_overview}
\end{figure}

\subsection{Setup}
In this initial phase, the $n$ clients acquire the necessary private and public parameters. These parameters are generated through a distributed mechanism using the $setup$ and $DKG$ algorithms in Algorithm \ref{alg:dve}. By the conclusion of this phase, a collective public key $pk_\alpha$ is generated, and each client securely generates a keypair, denoted as $(pk_i,sk_i)$. Additionally, the required trusted setup for zkSNARKs to generate the common reference string \((CRS)\) is generated. The \((CRS)\), which is crucial for proving and verifying zkSNARK proofs, is made accessible on the distributed ledger. 

\subsection{Submission}
In the submission phase, once a query is made by the collector, each client is expected to send a valid input value \(x_i\) to the aggregator. The clients pre-process the data and locally execute the \emph{DVE.Enc} algorithm (in Algorithm \ref{alg:dve}) on their input $x_i$. The output of this algorithm is $E(x_i)$ and proof $\Pi$. The encrypted input is subsequently transmitted to the aggregator and the proofs are submitted to the distributed ledger to allow for public verifiability. The encryption protects the privacy of the input data while the proof proves that the input is valid according to a predefined validation predicate $Valid(x)$. the VPAS protocol is compatible with any validation function that can be integrated into a general-purpose zkSNARK circuit as shown in Section \ref{sec:construction}.  

\subsection{Aggregation}
The aggregation phase which involves computing an aggregation function $f$ is asynchronous to user submissions and may be invoked by the aggregator at any time. Once a client submits an encrypted input, the aggregator utilizes the predefined validation predicate and accepts the client submission only if $Valid(x) = 1$ by running the \emph{DVE.VerifyEnc} in Algorithm \ref{alg:dve}. This verification process is essential to prevent incorrect or malicious data from impacting the final aggregated results. 

Additionally, it is essential to verify the aggregation step itself to prevent a malicious aggregator from sending malformed results, therefore, the aggregator must provide a proof of correct execution of the aggregation function. The proof is then submitted to the distributed ledger for public verifiability. To achieve this, the aggregator runs \emph{VA.Agg} in Algorithm \ref{alg:va}. The clients can then check that the received value is the correct aggregation result. 

\subsection{Release}
The Release phase represents the final step of the VPAS protocol. During this phase, the aggregated result of clients' encrypted inputs is made available to the collector. The VPAS protocol ensures that the released results remain unmodified during the release process. This is essential since the release phase in our setting involves additional re-encryption steps executed by the clients. Thus, similar to previous phases, this phase includes the validation step $Valid(y,\hat{y})$, where $y$ is the encrypted aggregate result and $\hat{y}$ is its re-encrypted counterpart. The validation in this phase ensures that the plaintext values within both $y$ and $\hat{y}$ are identical. Each client generates the re-encryption shares along with a proof by running the \emph{VRE.GenShare} algorithm in Algorithm \ref{alg:vre}. The proofs generated are publicly verifiable and published on the distributed ledger. The collector can then run the \emph{VRE.ReEnc} algorithm which checks each re-encryption share and generates $\hat{y}$, the aggregation result encrypted under the collector's public key. Finally, the collector can decrypt $\hat{y}$ using his secret key as described earlier in section \ref{subsec:phe-elgamal}.

\section{Security Analysis} \label{sec:security}
In this section, we briefly sketch the security proofs w.r.t. the security goals and the threat model stated earlier in section \ref{subsec:sec-goals}. Our protocol is constructed using a composition of three gadgets that were described in section \ref{sec:construction}. These gadgets are constructed using existing, peer-reviewed cryptographic primitives. Consequently, the VPAS protocol derives its security properties directly from these well-established primitives. Therefore, we refrain from a formal security proof, instead, we provide the following proof sketch on how VPAS archives data privacy, robustness, and computational correctness. 

\textbf{Data privacy.} 
As described in Section \ref{sec:protocol} the clients' encrypted data are never decrypted before being aggregated and re-encrypted with the collector’s public key. Therefore, as long as the distributed verifiable encryption (DVE) scheme described in Gadget 1 is secure, the VPAS protocol is secure against a malicious adversary $\mathcal{A}$ who can statically corrupt up to $n-1$ out of $n$ clients. The security of DVE follows that proven in \cite{lee2019saver} except for distributing the randomness during the generation of $pk_\alpha$. The randomness $t$ and $s$ are generated by all clients in the protocol as can be observed from the $DKG$ algorithm (Algorithm \ref{alg:dve}). Therefore, as long as at least one client is honest, the security of \emph{DVE} follows that in \cite{lee2019saver}. The ElGamal encryption used is secure under the discrete-log assumption, and the zkSNARK used inherits the correctness, soundness, and zero-knowledge property from the Groth16 scheme (proven in \cite{groth2016size}). Therefore, the adversary $\mathcal{A}$ cannot gain any information about the clients' input aside from that inferred from the proof. Additionally, subsequent operations on the ciphertext in both VA and VRE steps do not reveal any private information about the clients' input. VA aggregates the ciphertexts, and VRE re-encrypts the final results revealing nothing about the individual input or the secret keys used to encrypt it. The PoK proof used to verify re-encryption is also zero-knowledge (as shown in \cite{camenisch1997proof}) and only ensures the correct generation of re-encryption shares. To summarize, as long as one client is honest, an adversary $\mathcal{A}$ who controls the remaining clients, aggregator, and collector cannot break the confidentiality of the client's individual input data. 

\textbf{Result robustness.} 
VPAS ensures robustness since all clients' input values are verified using general-purpose zkSNARKs circuits. We rely on the Groth16 zkSNARKs scheme which has been proven to be complete, sound, and zero-knowledge \cite{lee2019saver, groth2016size}. By enforcing the generation of zkSNARK proofs on clients' input, the aggregated results are protected against malicious or malformed input. The robustness of the protocol holds even if the client and the aggregator collude since any data included in the results must have a proof submitted to the distributed ledger for public verifiability. 

\textbf{Computational correctness.} 
The VPAS protocol comprises three main computational steps: encryption, aggregation, and re-encryption. These steps utilize the three constructed gadgets and each gadget outputs not only the result of the computation performed but also a proof of correctness. Aside from validating the input, the \emph{DVE} proof ensures that the plaintexts were encrypted correctly as can be observed from Algorithm \ref{alg:dve}. The \emph{VA} proof ensures correct aggregation and the \emph{VRE} proof ensures that the re-encryption shares were generated correctly. These proofs ensure that the computations done during the execution of the VPAS protocol, both on the client and aggregator side, are correct. Additionally, all proofs generated by these gadgets are publicly verifiable ensuring that all parties, including those not directly participating in the protocol, can verify the correctness of the protocol's execution.

\section{Implementation and Experimental Evaluation} \label{sec:impl-eval}
\subsection{Implementation} \label{sec:impl}
We have implemented
\footnote{The source code will be made publicly available upon acceptance.} 
% \footnote{The source code link in Appendix \ref{code-link}.} 
the cryptographic gadgets for both Construction I (encryption-in-the-circuit) and Construction II. The in-circuit encryption is written primarily in Circom and we integrated the Circom circuits into Rust for evaluation. Our implementation leverages Circomlib, a library with common circuit implementations from iden3 \cite{iden3_circomlib} and we extend this library with optimizations that are summarized in appendix \ref{ecc-opt}. The code base for Construction II which includes the Exponential ElGamal encryption, DVE, VA, and VRE is written in Rust. We make use of the Arkworks framework \cite{arkworks2020}, a collection of Rust libraries for implementing ZKP proofs. Arkworks includes zkSNARKs implementation such as Groth16, which we use in this work. For efficiency, we extended the implementation to support proof aggregation using SNARKPACK \cite{gailly2022snarkpack}. Our resulting library is split into 2, one for construction I and the second for construction II. 

\begin{figure*}[ht]
\centering
\includegraphics[width=\textwidth]{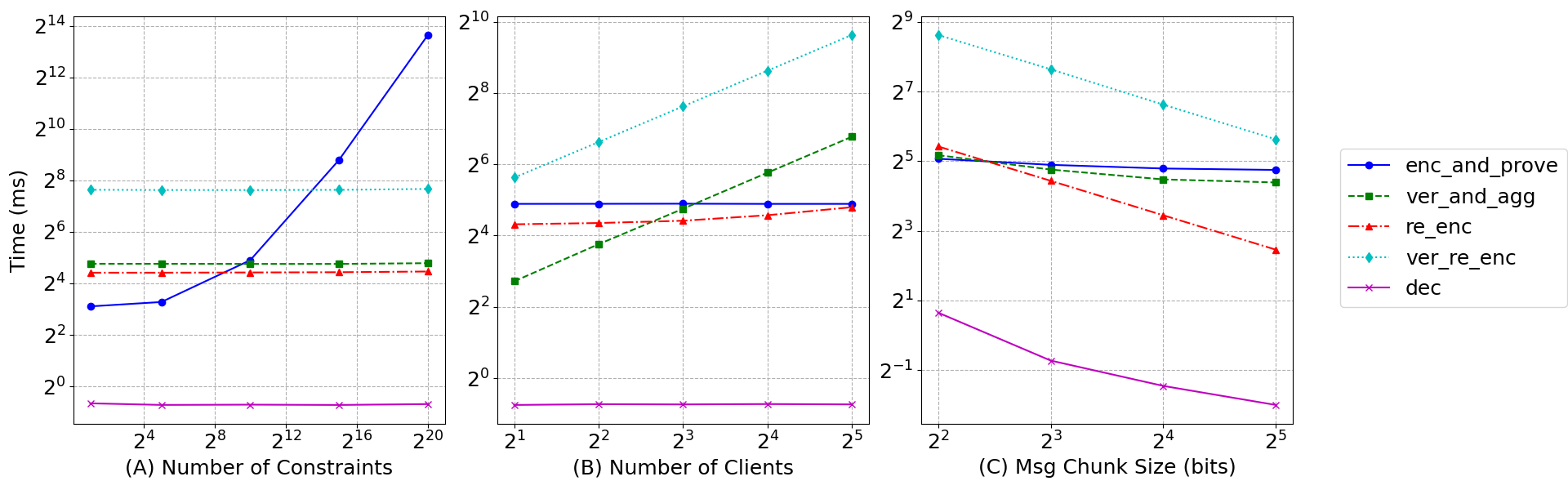}
\caption{Benchmarking results showcasing the impact of varying parameters: (A) Number of constraints with 8 clients and an 8-bit chunk size, (B) Number of clients with \(2^{10}\) constraints and an 8-bit chunk size, and (C) Message chunk sizes with \(2^{10}\) constraints and 8 clients.}
\label{fig:benchmarking_results}
\end{figure*}

\subsection{Experimental Evaluation}\label{sec:eval}
In this section, we present our experimental evaluation of the VPAS protocol, aiming to provide insights into its concrete efficiency. Our evaluation was conducted on a machine equipped with an Apple M2 Pro CPU and 16GB of RAM. We focused on two metrics: single-threaded runtime and communication cost. For each step of the VPAS protocol, we report the average results from ten trials. We adjust the following parameters to examine their impact on performance:
\begin{enumerate}
[leftmargin=0.5cm] 
    \item The number of rank-1 constraints, varying from $2$ to $2^{20}$.
    \item The size of the encryption message chunk size, ranging from $4$ bits to $32$ bits.
    \item The number of clients, increasing from $2$ to $2^{5}$.
\end{enumerate}
These metrics are pivotal for assessing the protocol's practical feasibility and scalability, particularly by examining the overhead induced by variations in these parameters.

\subsubsection{Setup Cost} 
The setup for zk-SNARKs, a one-time process, heavily relies on the circuit used for input validation and can be conducted when spare computational resources and bandwidth are available. Therefore, we omit the evaluation of the zkSNARKs setup. Similarly, the Distributed Key Generation (DKG) protocol, essential for generating the encryption key, is also a one-time procedure, unless new clients join or some existing ones drop out. The most expensive part of DKG is the communication between clients, which increases with the number of clients. For each client, generating the partial key $pk_i$ requires \(850\) ms, and computing $P_1$ takes \(280\) ms, assuming a message chunk size of 32-bits. The DKG protocol requires broadcasting both $pk_i$ and $P_1$, leading to \(2 \times N\) communication rounds, where $N$ represents the number of clients.

\subsubsection{Runtime Evaluation}
To demonstrate the scalability of VPAS, we evaluated each step of the protocol under varying numbers of constraints, clients, and message chunk sizes. The results are depicted in Figure \ref{fig:benchmarking_results}. These steps are executed by different parties involved in the protocol. Clients initiate the process by submitting their inputs through the \texttt{DVE.Enc} (\texttt{enc\_and\_prove}) step. This step is significantly impacted by the number of constraints, as the time required to generate the zk-SNARKs proof increases linearly with the number of constraints. Our findings indicate that, beyond $2^{15}$ constraints, this step becomes the dominant factor in runtime costs compared to other stages. Increasing the number of constraints does not significantly impact the other steps. However, increasing the number of clients degrades the performance of both the aggregation (\texttt{ver\_and\_agg}) and re-encryption (\texttt{ver\_re\_enc}) steps. This degradation is expected, as a higher number of clients results in an increased number of ciphertexts and proofs to be processed by the aggregator and collector in these two steps. Increasing the message chunk size improves performance across all steps. This improvement is primarily because smaller chunk sizes require more chunks to compose the entire message, thereby increasing the processing load in each step. We note that, generally, with larger chunk sizes, decryption— which involves finding the discrete logarithm as described in section \ref{subsec:phe-elgamal}—takes more time. However, in our implementation, we employ a preprocessing step to calculate the powers of $g$ and rely on the baby-step giant-step algorithm \cite{shanks1971class}, thereby making decryption efficient for chunk sizes up to 32-bit.

\begin{table}[htbp]
\begin{tabular}{|c|c|cc|cc|}
\hline
Input & Baseline & Const1 & Overhead & Const2 & Overhead \\
\hline
32 & 86.93 & 1087.83 & 12.51$\times$ & 108.33 & 1.25$\times$ \\
\hline
64 & 172.09 & 2195.54 & 12.76$\times$ & 215.89 & 1.25$\times$ \\
\hline
128 & 343.55 & 4364.28 & 12.70$\times$ & 431.86 & 1.26$\times$ \\
\hline
256 & 686.32 & 8734.28 & 12.73$\times$ & 863.17 & 1.26$\times$ \\
\hline
512 & 1372.09 & 17394.80 & 12.68$\times$ & 1732.3 & 1.26$\times$ \\
\hline
1024 & 2806.22 & 34871.89 & 12.43$\times$ & 3478.8 & 1.24$\times$ \\
\hline
\end{tabular}
\caption{Client Runtime Performance Benchmarks. The measurements for baseline, Construction I (Const1), and Construction II (Const2) were all measured in milliseconds.}
\label{tab:runtime_benchmarks_zkp}
\end{table}
\vspace{-0.5cm}

\subsubsection{Comparison with baseline}
We conducted an evaluation of Construction II in comparison to the encryption-in-the-circuit approach of Construction I. We also compared it against a baseline approach that provides privacy without input validation, i.e. assuming that both the clients and the aggregator are honest-but-curious. We measured the time it takes for clients to perform encryption using both constructions and the baseline. We used the BN254 \cite{barreto2005pairing} and set the message chunk size to be 32-bits. It is important to note that in both constructions, the encryption algorithm generates not only the ciphertext but also a proof. The runtime increases linearly with the number of constraints. We used a circuit with a single constraint which is sufficient to show the difference in overhead between the two constructions. Table \ref{tab:runtime_benchmarks_zkp} presents the average results from ten trials with a varying number of inputs. The baseline illustrates the cost associated with achieving privacy alone. Construction I, which employs encryption within the circuit, incurs a significantly high cost primarily due to the proof generation, resulting in approximately a $12.6\times$ slowdown. Conversely, Construction II, while still slower compared to the baseline, proves to be substantially more efficient than Construction I, with the client-side cost averaging a $1.25\times$ increase.

\subsubsection{Communication Cost}
Analyzing the size of data exchanged provides insights into the efficiency of the VPAS protocol in terms of data communication. We measured the sizes of parameters exchanged between parties, as shown in Table \ref{tab:size_benchmarks_all_chunks}. Furthermore, we evaluated the communication costs for each protocol step, as illustrated in Table \ref{table:communication_overhead}. The Key Generation step incurs costs of \(N \times |pk_i| + N \times |P_1|\), which depend on the number of clients (\(N\)), the size of the public key (\(|pk|\)), and the size of \(|P_1|\). The communication cost of the Distributed Verifiable Encryption (DVE) step increases with the number of clients, involving the ciphertext size (\(|CT|\)) and the size of the DVE proof (\(|\pi_{DVE}|\)). The Aggregation step require one round of communication, where the aggregator broadcasts the aggregated result (\(|\hat{CT}|\)). Finally, the Verifiable Re-encryption (VRE) step, formulated as \(N \times (|w_1| + |w_2|) + N \times |\pi_{VRE}|\), is influenced by the number of clients (as each client participates in the re-encryption), the size of the re-encryption shares (\( |w_1|, |w_2| \)), and the size of the VRE proof $\pi_{VRE}$.

\begin{table}[htbp]
\centering
\small
\begin{tabular}{|c|c|c|c|c|}
\hline
\textbf{Chunk Size} & \textbf{4-bit} & \textbf{8-bit} & \textbf{16-bit} & \textbf{32-bit} \\
\hline
\textbf{$pk_i$} & 12504 & 6360 & 3288 & 1752 \\
\hline
\textbf{$P_1$} & 48 & 48 & 48 & 48 \\
\hline
\textbf{$pk_\alpha$} & 12552 & 6408 & 3336 & 1800 \\
\hline
\textbf{$CT$} & 3176 & 1640 & 872 & 488 \\
\hline
\textbf{$W_1$} & 48 & 48 & 48 & 48 \\
\hline
\textbf{$W_2$} & 3080 & 1544 & 776 & 392 \\
\hline
\textbf{$\pi_{DVE}$} & 192 & 192 & 192 & 192 \\
\hline
\textbf{$\pi_{VRE}$} & 10248 & 5128 & 2568 & 1288 \\
\hline
\end{tabular}
\caption{Parameters Sizes}
\label{tab:size_benchmarks_all_chunks}
\end{table}
\vspace{-0.8cm}
\begin{table}[h!]
\centering
\begin{tabular}{|l|l|}
\hline
\textbf{Protocol} & \textbf{Communication Cost} \\
\hline
$DKG$ & \( N \times |pk_i| + N \times |P_1| \) \\
\hline
$DVE$ & \( N \times |CT| + N \times |\pi_{DVE}| \) \\
\hline
$VA$ & \( |CT| \) \\
\hline
$VRE$ & \( N \times (|w_1| + |w_2|) + N \times |\pi_{VRE}| \) \\
\hline
\end{tabular}
\caption{Communication Cost}
\label{table:communication_overhead}
\end{table}
\vspace{-0.5cm}

\subsubsection{Case Study: GWAS Aggregate Statistics} \label{subsec:gwas-stat}
In Section \ref{subsec:app-scenario} we discussed the beacon framework, designed to facilitate a scenario in which a researcher seeks aggregate results from various genomic data custodians. The data custodians wish to protect their private input and only release the aggregate results, while the researcher wishes to ensure the authenticity and correctness of the aggregation process. Here, we discuss how VPAS performs in such a setting.

In our setup, $N$ genomic data custodians collaborate, agreeing to conduct a statistical analysis and to release the results to the researcher. In our evaluation, we consider calculating the Minor Allele Frequency (MAF) and Chi-squared ($\chi^2$) which are two common statistical tests used in Genome-Wide Association Studies (GWAS). During the initialization phase, the clients collectively generate the encryption key $pk_\alpha$. Furthermore, the parties agree on the zkSNARKs circuits. These circuits define the constraints for input validation. In our case, we employed a circuit designed to ensure correct encoding and verify the authenticity of input data through inclusion checks against public Merkle commitments. This circuit contains around 6000 constraints. Details on the input validation process are further elaborated in Appendix \ref{validation}. The public parameters, including $pk_\alpha$, $CRS$, and Merkle commitments, are published on a distributed ledger. The encoding we used maps the input SNP values to 8 32-bit chunks. Details on the encoding and aggregation functions we used in the case study are described in Appendix \ref{encoding}. Figure \ref{fig:gwas_results} presents the results of our evaluation based on the aforementioned setup. We vary the number of input aggregated to show the scalability of the protocol. The results suggest that 1000 input data can be processed through the VPAS in approximately 8 seconds. The evaluation was done on a standard PC, meaning that this time could be further reduced by using higher-performance machines. As illustrated in Figure \ref{fig:gwas_results}, the runtime for the aggregator and collector is higher, primarily due to the proof verification carried out during both the aggregation and re-encryption phases. This client runtime is low which can be attributed to the fact that proof generation is executed in parallel by each client on separate machines.

\begin{figure}[htbp]
\centering
\includegraphics[width=0.33\textwidth]{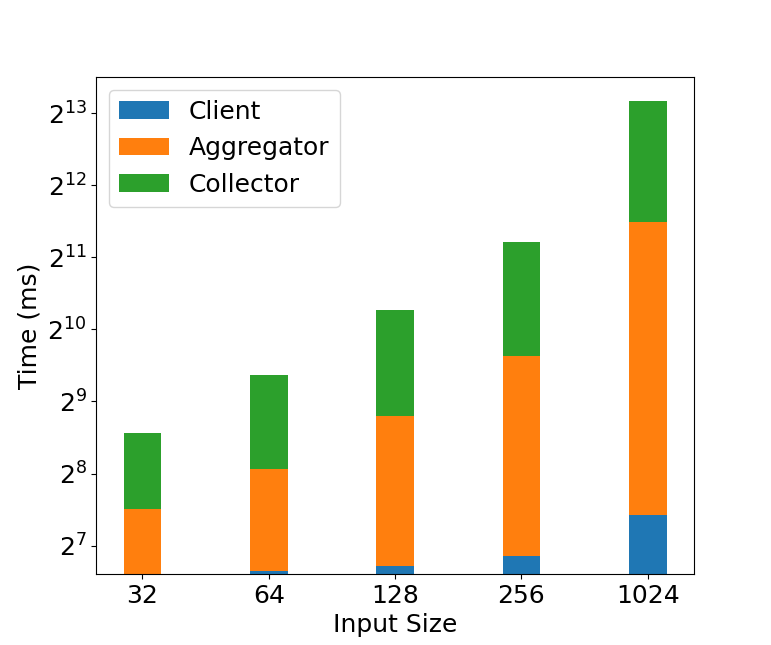}
\caption{GWAS benchmark results with varying input.}
\label{fig:gwas_results}
\end{figure}
\vspace{-0.2cm}

\section{Conclusion}\label{sec:conclusion}
In this paper, we introduced VPAS, a protocol designed to enable privacy-preserving aggregation of distributed datasets. By leveraging a combination of homomorphic encryption (HE) and Zero-Knowledge Proofs (ZKPs), VPAS addresses the concerns of data privacy, input validation, computational correctness, and public verifiability in the context of secure data aggregation. These attributes make VPAS particularly suitable for applications requiring stringent data privacy and verifiability, such as in healthcare, where data confidentiality and integrity are critical. Our results demonstrate the practicality and efficiency of VPAS, particularly highlighting its lower computational overhead compared to simply using conventional zkSNARKs. This is achieved through the construction of Distributed Verifiable Encryption (DVE), Verifiable Aggregation (VA), and Verifiable Re-Encryption (VRE), a set of novel algorithms introduced in this work. Future studies could explore alternative zkSNARKs that offer different trade-offs, such as those with a universal setup or transparent zkSNARKs that eliminate the need for a trusted setup. Additionally, another avenue of research is to explore enhancing the utility of the aggregation function with verifiable fully homomorphic encryption.

%% ref in new page
% \newpage
\bibliographystyle{ACM-Reference-Format}
\bibliography{ref}

\clearpage

\appendix

\section{Optimized Implementation of In-Circuit Homomorphic Encryption} \label{ecc-opt}
In this section, we discuss our techniques for optimizing the in-circuit encryption of exponential Elgamal encryption. Here, we focus on lowering the number of non-linear constraints of the circuits which is an important metric for the efficiency of a circuit. A non-linear constraint is produced when a computation involving the product of two input signals is made. For example, if $k$ and $l$ are input signals to a circuit then computing the product $k \cdot l$ in the circuit will add a non-linear constraint. If instead $l$ is a fixed value defined in the circuit then computing the product $k \cdot l$ would not produce a non-linear constraint. On this note, to improve on the efficiency of the often-used in-circuit homomorphic encryption program written by iden3 \cite{iden3_circomlib} we have constructed such a circuit in a way that reduces the number of non-linear constraints by around 600 from 3859 to 3260.

In Table \ref{tab:hom_enc_table}, we state the number of non-linear constraints for both the baseline (iden3's) implementation and our optimized implementation of in-circuit encryption.

\begin{table}[htbp]
\centering
\begin{tabular}{ccc}
    \toprule
    Hom. Enc. & Sub-function & Non-linear \\
    & & constraints \\
    \midrule
    \multirow{5}{*}{baseline} & \textbf{Total} & \textbf{3859} \\
    & \texttt{Bitify (x2)} & 506 \\
    & \texttt{Scal. Mult. Any} & 2301 \\
    & \texttt{Scal. Mult. Gen. (x2)} & 1046 \\
    & \texttt{Tw. Edw. Addition} & 6 \\
    \midrule
    \multirow{4}{*}{improved} & \textbf{Total} & \textbf{3260} \\
    & \texttt{Scal. Mult. Any} & 1757 \\
    & \texttt{Scal. Mult. Gen. (x2)} & 1500 \\
    & \texttt{Mont. Addition} & 3 \\
    \bottomrule
\end{tabular}
\caption{Number of non-linear constraints produced by in-circuit homomorphic encryption}
\label{tab:hom_enc_table}
\end{table}

Both implementations include circuits for scalar multiplication of arbitrary on-curve points (\texttt{Scal. Mult. Any}) as well as scalar multiplication of a chosen generator (\texttt{Scal. Mult. Gen.}). As can be seen in Table \ref{tab:hom_enc_table}, the improvement is mostly due to a far less costly implementation of scalar multiplication of arbitrary on-curve points as well as the lack of a need for iden3's \texttt{Bitify} circuit which is used in their implementation. We use the Baby Jubjub curve whose base field matches the scalar field of BN254 \cite{whitehat2020baby}. This EC choice allows for efficient cryptographic operations.

In benchmarking the improved implementation's homomorphic encryption program, we measure the time taken to complete two stages of the proof, proof setup and proof generation, given multiple inputs. The average reduction in run time for the proof setup is 33\%. For the proof generation phase, there is less reduction with a 17.1\% average reduction. 

The implementation can be improved further by utilizing the baseline implementation's circuit for scalar multiplication of a chosen generator along with a circuit for mapping points from Twisted Edwards form to Montgomery form and vice versa. The use of the former would reduce the number of non-linear constraints produced by our homomorphic encryption program by around 450 while the use of a program that maps points on our elliptic curve from Twisted Edwards form to Montgomery form and vice versa would add around 10 non-linear constraints. As such, we estimate that this improvement would reduce the number of non-linear constraints of our homomorphic encryption by around 440 to an overall cost of around 2820 non-linear constraints. Compared to the baseline implementation, this would yield a significant reduction of 1040 non-linear constraints.

\subsection{Overview of Optimization Techniques}

In the baseline implementation, iden3's circuit implementations for scalar multiplication are used: \texttt{escalarmulfix} for scalar multiplication of a pre-chosen generator and \texttt{escalarmulany} for scalar multiplication of an arbitrary point on the curve. These operations incur costs of $523$ and $2301$ non-linear constraints, respectively. Both \texttt{escalarmulfix} and \texttt{escalarmulany} employ the sliding-window method, an efficient technique for computing scalar multiples on elliptic curves. It is important to note that the scalar input to these circuits must be provided in binary form. Consequently, a conversion circuit is required to transform the integer scalar into its binary equivalent. In the context of iden3, this is facilitated by the \texttt{Bitify} circuit, as shown in Table \ref{tab:hom_enc_table}, contributing an additional $253$ non-linear constraints per invocation.

\subsubsection{Double-and-add for arbitrary on-curve points}
To improve upon the unusually high number of non-linear constraints produced by \texttt{escalarmulany}, our improved implementation relies on the double-and-add method of scalar multiplication. Despite double-and-add being less efficient than the sliding-window method, the cost of scalar multiplication of arbitrary on-curve points on Baby Jubjub in the new implementation is heavily reduced.

The following is an overview of how double-and-add for arbitrary on-curve points was implemented. Note that this is not an in-depth line-by-line assessment of how the circuit is written but instead acts as a summary of what is written, the underlying challenges faced while writing this in Circom, and how these challenges were resolved.

Traditionally, the double-and-add method of elliptic curve scalar multiplication is done in a way similar to the following. Given the scalar $k$ and base point $P$ as input, one computes $[k]P$ as is described in Algorithm \ref{alg:regular_double_and_add}.

\begin{algorithm}[htbp]
\caption{Traditional double-and-add for scalar multiplication on an elliptic curve $E/\mathbf{F}_p$}
\label{alg:regular_double_and_add}
\begin{algorithmic}[1]
\REQUIRE An integer $k>0$ and a point $P\in E(\mathbf{F}_p)$
\ENSURE $R=[k]P\in E(\mathbf{F}_p)$
\STATE $R\gets\mathcal{O}$
\STATE $firstBit\gets k \% 2$
\WHILE{$k > 0$}
    \STATE $bit\gets k \% 2$
    \IF{bit $= 1$}
        \STATE $R\gets R+P$
    \ENDIF
    \STATE $P\gets P+P$
    \STATE $k\gets k>>1$
\ENDWHILE
\RETURN $R$
\end{algorithmic}
\end{algorithm}

In implementing the double-and-add method in Circom, one must take into account that input signals, i.e. the input scalar $k$, and its modified state can not be used explicitly as a conditional. As such, implementing double-and-add is not as simple as in the brief pseudocode written in Algorithm \ref{alg:regular_double_and_add}. To get around this, we instead implement something similar to Algorithm \ref{alg:condition_free_double_and_add}.

\begin{algorithm}[htbp]
\caption{Condition-free double-and-add for scalar multiplication on an elliptic curve $E/\mathbf{F}_p$}
\label{alg:condition_free_double_and_add}
\begin{algorithmic}[1]
\REQUIRE An integer $k>0$ and a point $P\in E(\mathbf{F}_p)$
\ENSURE $R=[k]P\in E(\mathbf{F}_p)$
\STATE $R\gets P$
\STATE $firstBit\gets k \% 2$
\WHILE{$k > 0$}
    \STATE $bit\gets k \% 2$
    \STATE $R\gets[bit](R+P)+[1-bit](R+\mathcal{O})$
    \STATE $P\gets P+P$
    \STATE $k\gets k>>1$
\ENDWHILE
\STATE $R\gets R-[1-firstBit]P$
\RETURN $R$
\end{algorithmic}
\end{algorithm}

The current bit of $k$ is given by $k\%2$ and so if it is 0 then the additive identity $\mathcal{O}$ is added to the returned point $R$, i.e. $R$ remains the same as before adding, and if the current bit is $1$ then $P$ is added to $R$, as in regular double-and-add. In doing this, the input scalar $k$ is not used explicitly as a conditional.

The second consideration to make when implementing this in Circom is that in step 7 of Algorithm \ref{alg:condition_free_double_and_add},
$$R\gets[bit](R + P) + [1 - bit](R + \mathcal{O}),$$
the sign $+$ denotes Montgomery elliptic curve addition. As such, adding $\mathcal{O}$ to the returned point $R$ requires adding by a point that does not change $R$. For a curve in Twisted Edwards form, this can be achieved simply by adding the additive identity $(0, 1)$ to $R$ but in Montgomery form, Baby Jubjub does not have an additive identity, which poses the following challenge: given a point $R=(x_3, y_3)$ on an elliptic curve in Montgomery form, what points $P=(x_1,y_1)$ and $Q=(x_2,y_2)$ satisfy $P+Q=R$?

To address this, note the following. Given points $P=(x_1, y_1)$ and $Q=(x_2, y_2)$ on an elliptic curve in its Montgomery form with coefficients $A$ and $B$, one computes the sum $P+Q=(x_3, y_3)$ as in Algorithm \ref{alg:mont_addition}.

\begin{algorithm}[htbp]
\caption{Point addition on a Montgomery elliptic curve $E/\mathbf{F}_p$}
\label{alg:mont_addition}
\begin{algorithmic}[1]
\REQUIRE Points $P=(x_1, y_1)$, $Q=(x_2, y_2)\in E(\mathbf{F}_p)$
\ENSURE $P+Q=(x_3, y_3)\in E(\mathbf{F}_p)$
\STATE $\lambda\gets(y_2 - y_1)(x_2 - x_1)^{-1}$
\STATE $x_3\gets B\lambda^2 - A - x_1 - x_2$
\STATE $y_3\gets\lambda(x_1 - x_3) - y_1$
\RETURN $(x_3, y_3)$
\end{algorithmic}
\end{algorithm}

On this note, given a point $R=(x,y)$ and taking $P=(-x-A, -y)$ and $Q=(0, -y)$, Algorithm \ref{alg:mont_addition} yields $(x_3, y_3) = (x, y)$ and so $P+Q=R$. As such, in implementing double-and-add in Circom, if the current bit of $k$ is 0, we instead set the returned point $R$ to be the sum of the points $(-x-A, -y)$ and $(0, -y)$ where $x$ and $y$ are the coordinates of $R$ before addition. With this in mind, our approach to double-and-add can be summarised by Algorithm \ref{alg:circom-tailored_double_and_add}.

\begin{algorithm}[htbp]
\caption{Circom-tailored double-and-add for scalar multiplication on an elliptic curve $E/\mathbf{F}_p$}
\label{alg:circom-tailored_double_and_add}
\begin{algorithmic}[1]
\REQUIRE An integer $k>0$ and a point $P\in E(\mathbf{F}_p)$
\ENSURE $R=[k]P\in E(\mathbf{F}_p)$
\STATE $R\gets P$
\STATE $firstBit\gets k \% 2$
\WHILE{$k > 0$}
    \STATE $bit\gets k \% 2$
    \STATE $R\gets[bit](R+P)+[1-bit]((-R_x-A, -R_y) + (0, -R_y))$
    \STATE $P\gets P+P$
    \STATE $k\gets k>>1$
\ENDWHILE
\STATE $R\gets R-[1-firstBit]P$
\RETURN $R$
\end{algorithmic}
\end{algorithm}

In summary, Algorithm \ref{alg:circom-tailored_double_and_add} is simply an implementation of double-and-add where `doubling' always occurs and `adding' occurs only when the leading bit of the (modified) scalar $k$ is 1. This relies on no conditionals dependent on the input scalar $k$ all while operating on the curve \textit{only} in Montgomery form.

\subsubsection{Double-and-add for a pre-chosen generator}

For scalar multiplication of a pre-chosen generator, an almost identical approach is taken. The only change is that the set of doubles of the pre-chosen generator is pre-computed and stored in a separate circuit. In doing this, one heavily reduces the number of non-linear constraints required as no doubling is needed. It's advised that the pre-computation of these doubles is done in a different language, as computing these doubles in Circom itself and formatting the corresponding Circom circuit manually would be heavily time-consuming.

\section{Proofs of Knowledge of the Discrete Logarithm} \label{pok}
A proof of knowledge allows a prover to convince a verier that she knows a solution to a problem that is hard to solve such as finding the discrete logarithm. In this case, given $g^x$ the prover proves knowledge of $x$ without revealing the value $x$. In our work, we use this proof of knowledge to verify that two discrete logarithms satisfy a linear equation. As described in \cite{camenisch1997proof}, given two discrete logarithms $y_1 = g_1^{x_1}, y_1 = g_2^{x_2}$, the prover proves that they satisfy the equation:
\[
a_1^{x_1} \cdot a_2^{x_2} = b
\]

Where $a_1, a_2, b$ are publicly known and $x_1, x_2$ are kept private. We omit the details of this proof system and refer the reader to \cite{camenisch1997proof}. 
\subsection{Proving correctness of re-encryption shares}
To prove the correctness of the re-encryption shares ($w_1,w_2$) and given that $pk = (X_0, X, Y, Z, P_2)$, the prover uses the above proof of knowledge where the two discrete logarithms are $y_1 = X, y_2 = w_1$, the public values are $a_1 = -C_1, a_2 = pk_\beta, b = w_2$, and the private values are $x_1 = s, x_2 = z$. Essentially the verifier checks the following equality:
\[
-C_1^{s} \cdot pk_\beta^{z} = w_2
\]
Essentially the verifier checks that the $s$ and $z$ values used in ($pk, w_1$) are equal to the ones used to generate $w_2$.

\section{Case Study: GWAS Aggregate Statistics}
\label{gwas}

\subsection{GWAS Background}
The human genome contains 3 billion nucleotide pairs, and  99\% of it is shared among humans. The genetic variation accounts for only 1\%, and they indicate unique biological characteristics including genetic predispositions to diseases. The majority of these variations are called Single Nucleotide Polymorphisms (SNPs). Within each SNP, the major allele is the most commonly found nucleotide in a population, and the minor allele is the least common. Some of the most notable examples of successful GWAS studies include the identification of genetic variations associated with diseases such as type 2 diabetes, Crohn's disease, and various forms of cancer.

During a typical GWAS, researchers use a number of statistical algorithms. We consider the Minor Allele Frequency (MAF) and the $\chi^2$ test. 

Table \ref{tbl:dataset} shows a typical (raw) dataset and Table \ref{tbl:snp} is a 3$\times$2 contingency table for each SNP which is commonly used when performing GWAS. \( N^{pop}_{i} \) is the count of the alleles \( i \in \{AA, Aa, aa\} \) in the case/control population \( pop \in \{case, control\} \). \( N^{case} \) and \( N^{control} \) are the size of the case and the control population, respectively.

\begin{table}
        \small
        \begin{tabular}{ |c|c|c|c|c|c| }
            \cline{2-6}
             \multicolumn{1}{c|}{} & \textbf{$SNP_1$} & \textbf{$SNP_2$} & \textbf{...} & \textbf{$SNP_m$} & \textbf{Population} \\ 
            \cline{2-6} \hline
            $gen_1$ & AG & AG & ... & AA & case\\ 
            \hline
            $gen_2$ & CT & CC & ... & CT & control\\ 
            \hline
            ... & ... & ... & ... & ... & ... \\ 
            \hline
            $gen_n$ & AA & AT & ... & TT & control\\ 
            \hline 
        \end{tabular}
    \caption{Raw Data for GWAS \label{tbl:dataset}}
   \reducespaceThalf
\end{table}

%\vspace{-0.3cm}

\begin{table}
        % \small
        \begin{tabular}{ c|c c|c }
            %\cline{2-4}
             \multicolumn{1}{c|}{} & \textbf{Case} & \textbf{Control} & \textbf{Total} \\ 
            \cline{2-4} \hline
            AA & $N_{AA}^{case}$ & $N_{AA}^{control}$ & $N_{AA}$\\ 
            %\hline
            Aa & $N_{Aa}^{case}$ & $N_{Aa}^{control}$ & $N_{Aa}$\\ 
            %\hline
            aa & $N_{aa}^{case}$ & $N_{aa}^{control}$ & $N_{aa}$ \\ 
            \hline
            \textbf{Total} & $N^{case}$ & $N^{control}$ & $N^{Tot}$ \\ 
            %\hline 
        \end{tabular}
    \caption{Contingency Table for a Single SNP \label{tbl:snp}}
    \reducespaceT
\end{table}
The Minor Allele Frequency (MAF), which, as the name suggests,
is the frequency at which the second most common allele occurs
in a given population. Both MAF and the $\chi^2$ test allow researchers to identify the top alleles whose frequencies differ the most between the case and control populations. To compute these statistics, the allele counts are calculated first using the values in Table \ref{tbl:snp} for the case and control population as follows:

\begin{equation}\label{eq:maf-maj}
    N_A^{pop} = N^{pop}_{Aa}+2N^{pop}_{AA}
\end{equation}

\begin{equation}\label{eq:maf-min}
    N_a^{pop} = N^{pop}_{Aa}+2N^{pop}_{aa}
\end{equation}

MAF is then calculated by the following formula:
\begin{equation}\label{eq:maf-equation}
    {MAF}^{pop} = \frac{ MIN (N^{pop}_{A}, N^{pop}_{a})}{2N^{pop}}
\end{equation}

The $\chi^2$ test is computed as shown below:
\begin{equation}\label{eq:chi-eq}
\chi^2 = \sum_{i \in \{A,a\}} \frac{(N_i^{case} - N_i^{control})^2}{N_i^{control}}.
\end{equation}

From the $\chi^2$ value, the p-value of each SNP can be computed. The p-value indicates that the variant might be significant if it is lower than a given threshold (e.g., $10^{-8}$) \cite{barsh2012guidelines}. 

\subsection{Encoding}\label{encoding}
In this section, we provide the details of how we encoded the SNP data in the case study. Encoding is a common technique used in aggregate statistics protocols and can be seen in prior work \cite{froelicher2020drynx,corrigan2017prio}. We start with the SNP encoding, and then in the next subsections, we describe the aggregation function for calculating MAF and the $\chi^2$ test. 

SNP data are organized in a table in the form of Table \ref{tbl:dataset}. Assuming that SNPs are biallelic, the possible combination for the two alleles within the SNP are (AA, Aa, aa). Each allele is one of the possible nucleotide bases A, C, T, and G. We use an encoding approach similar to that in \cite{lauter2014private}. Encoding each allele pair for SNP $i$ and gen $j$ of Table \ref{tbl:dataset} can be done by mapping the pairs using the following rule:

\[
\begin{aligned}
AA:\quad c_0^{(i,j)} \leftarrow 1,\quad c_1^{(i,j)} \leftarrow 0,\quad c_2^{(i,j)} \leftarrow 0 \\
Aa:\quad c_0^{(i,j)} \leftarrow 0,\quad c_1^{(i,j)} \leftarrow 1,\quad c_2^{(i,j)} \leftarrow 0 \\
aa:\quad c_0^{(i,j)} \leftarrow 0,\quad c_1^{(i,j)} \leftarrow 0,\quad c_2^{(i,j)} \leftarrow 1
\end{aligned}
\]

Then we consider each $N_{AA}^i, N_{Aa}^i, N_{aa}^i$ for SNP $i$ as sets:
\[
\begin{aligned}
N_{AA}^i = \{c_0^{(i,j)}\}_{j=0}^n \\
N_{Aa}^i = \{c_1^{(i,j)}\}_{j=0}^n \\
N_{aa}^i = \{c_2^{(i,j)}\}_{j=0}^n 
\end{aligned}
\]

\subsubsection{MAF Aggregation}
While we are not able to obtain the allele frequencies due to the required division, we can obtain the counts using homomorphic additions. Indeed, if the SNPs are encoded as described earlier, then we can define four variables for each population, $N_{AA}^{pop}$, $N_{Aa}^{pop}$, $N_{aa}^{pop}$, and the number of available genotypes $N^{pop}$. Each client ($U_i$) calculates, encrypts, and submits these variables to the aggregators. Once these variables are received, the aggregator can produce the sum of these shares as shown in Table \ref{tab:maf-shares}. Then, the aggregator 
calculates $N_{A}^{pop}, N_{a}^{pop}$ as shown previously in Equation \ref{eq:maf-maj} and \ref{eq:maf-min}. The collector can then use these values to calculate MAF as shown in equations \ref{eq:maf-equation}.

\begin{table}[t]
\centering
\renewcommand{\arraystretch}{1.5}
\begin{tabular}{|c|c|c|c|c|}
\hline
\textbf{Client} & \multicolumn{4}{c|}{ \textbf{SNP Shares}} \\
\hline
$U_1$ & $N_{AA}^1$ & $N_{Aa}^1$ & $N_{aa}^1$ & $N^1$ \\
\hline
$U_2$ & $N_{AA}^2$ & $N_{Aa}^2$ & $N_{aa}^2$ & $N^2$ \\
\hline
$\vdots$ & $\vdots$ & $\vdots$ & $\vdots$ & $\vdots$ \\
\hline
$U_k$ & $N_{AA}^k$ & $N_{Aa}^k$ & $N_{aa}^k$ & $N^k$ \\
\hline
SUM & $N_{AA}$ & $N_{Aa}$ & $N_{aa}$ & $N$ \\
\hline
\end{tabular}
\caption{Shares of Data for Alleles Frequencies}
\label{tab:maf-shares}
\end{table}

\subsubsection{$\chi^2$ Aggregation}
Given the calculated allele counts, $\chi^2$ can be computed as follows: 

\begin{equation}\label{eq:chi-eq-exp}
\chi^2 = \frac{(N_A^{case} - N_A^{control})^2}{N_A^{control}} + \frac{(N_a^{case} - N_a^{control})^2}{N_a^{control}}
\end{equation}

This is an expanded version of Equation \ref{eq:chi-eq}. Since we are only able to perform aggregation through homomorphic addition, subtraction, or scalar multiplication, the aggregator computes the two additional values $\beta_A$ and $\beta_a$ which equal to:
\[
\begin{aligned}
\beta_A &= N_A^{case} - N_A^{control}, \\
\beta_a &= N_a^{case} - N_a^{control}, 
\end{aligned}
\]

The $X^2$ value can then be computed by the collector as

\begin{equation}\label{eq:chi-eq-beta}
\chi^2 = \frac{(\beta_A)^2}{N_A^{control}} + \frac{(\beta_a)^2}{N_a^{control}}
\end{equation}

\subsubsection{Encryption of encoded data}
After encoding, input data needs to be packed as messages for encryption. The message size can be arbitrary, however, assuming the message size is limited to 256-bit, and given that 4 terms are required for the case shares, and 4 terms for the control shares (as defined previously), we fix the message chunk size to  32-bits. Each chunk would represent one term: 
\[
M = \{M^{case}, M^{control}\}
\]
which expands to
\[
M^{case} = \{N_{AA}^{case},N_{Aa}^{case},N_{aa}^{case},N^{case}\}
\]
\[
M^{control} = \{N_{AA}^{control},N_{Aa}^{control},N_{aa}^{control},N^{control}\}
\]
The chunk size determines the ciphertext size and decryption time. A larger chunk size can fit a larger message space. However, as a trade-off, it increases the decryption time due to the increased computation of discrete log search. For this case study, we considered a 32-bit chunk which puts an upper limit of $2^{32}$ on the size of each population. 

\subsection{Input data validation} \label{validation}
The $Valid(x)$ algorithm ensures that each input $x$ is a valid encoded value of an SNP value $x \in \{0,1,2\}$. Specifically, the algorithm checks the following:

\[
Valid(x): (0 \geq x \geq 2) \rightarrow  \{0,1\}
\]

Additionally, the $Valid(x)$ checks if the input $x$ is part of the committed data. The Merkle commitments to the underlying data are important for the integrity of the results. In the event of a future dispute concerning the correctness of the overall statistical information, an investigator can request the providers to open their Merkle commitments and any provider that committed to incorrect data will be identified. The implemented $Valid(x)$ check that $x$ is included in the Merkle commitment given $R$ is the root of the Merkle tree and $\pi$ is the proof of inclusion as illustrated in the following:

\begin{itemize}[leftmargin=1em]
    \item $\text{Insert}(x) \rightarrow MT'$: adds the value $x$ into the next free leaf in the tree $MT$ and outputs the modified tree $MT'$.

    \item $\text{getRoot}() \rightarrow R$: returns the current root of the tree $R$.

    \item $\text{Prove}(x) \rightarrow  \pi$: given the value $x$, outputs a membership proof $\pi$ such that $H(x) \in MT$.

    \item $\text{Verify}(x,R,\pi) \rightarrow  \{0,1\}$: on inputs root R, value $x$ and proof $\pi$ outputs 1 if the value $x$ is in the Merkle tree, and 0 otherwise. 
\end{itemize}

In our case study, the $Valid(x)$ function incorporates the two previous checks and links the proofs to the ciphertexts generated in the previous step. In Construction I, the checks are written in the same circuit as the encryption, whereas in Construction II, the encryption is linked through the verifiable encryption algorithm described in section \ref{subsec:const2}.

% \section{Code} \label{code-link}
% Our open-source implementation is available at: \\ https://anonymous.4open.science/r/vpas-BE52/

\end{document}